# A Deep Network for Explainable Prediction of Non-Imaging Phenotypes using Anatomical Multi-View Data

Yuxiang Wei[1,2], Yuqian Chen[2,3], Tengfei Xue[2,3], Leo Zekelman[2] , Nikos Makris[2] , Yogesh Rathi[2] , Weidong Cai[3] , Fan Zhang[1,2] , and Lauren J. O'Donnell[2*]

[1] University of Electronic Science and Technology of China, Chengdu, China
[2] Harvard Medical School, Boston, USA
[3] University of Sydney, Sydney, Australia

**Abstract.** Large datasets often contain multiple distinct feature sets, or views, that offer complementary information that can be exploited by multi-view learning methods to improve results. We investigate anatomical-multi-view data, where each brain anatomical structure is described with multiple feature sets. In particular, we focus on sets of white matter microstructure and connectivity features from diffusion MRI, as well as sets of gray matter area and thickness features from structural MRI. We investigate machine learning methodology that applies multi-view approaches to improve the prediction of non-imaging phenotypes, including demographics (age), motor (strength), and cognition (picture vocabulary). We present an explainable multi-view network (EMV-Net) that can use different anatomical views to improve prediction performance. In this network, each individual anatomical view is processed by a view-specific feature extractor and the extracted information from each view is fused using a learnable weight. This is followed by a wavelet-transform-based module to obtain complementary information across views which is then applied to calibrate the view-specific information. Additionally, the calibrator produces an attention-based calibration score to indicate anatomical structures' importance for interpretation. In the experiments, we demonstrate that the proposed EMV-Net significantly outperforms several state-of-the-art methods designed for non-imaging phenotype prediction based on the Human Connectome Project (HCP) dataset with 1065 individuals. Specifically, our method reduces age prediction MAE for at least 2.4 years and elevates the correlation coefficient on predicting the other two phenotypes for at least 0.13. Our interpretation results show that for different views, fractional anisotropy of white matter diffusion measures and the surface thickness of gray matter measures are generally more important.

## 1      Introduction

The brain's anatomical structures are crucial in neurological function and neurodevelopment. Magnetic resonance imaging (MRI) enables quantitative analysis of the brain's structural properties. Structural MRI can measure the macroscopic morphome-try of the cortical and subcortical structures in gray matter (GM), while diffusion MRI (dMRI) tractography [1] can assess white matter (WM) connectivity and is used to extract quantitative microstructure measures such as fractional anisotropy (FA) and mean diffusivity (MD) [2]. To uncover links between brain structure and non-imaging phenotypes including demographics or behavioral traits, a popular avenue of research is machine learning (ML) for the prediction of non-imaging phenotypes. For example, ML models including ridge regression [3] and a lightweight deep-learning model named SFCN (Simple Fully Convolutional Network) [4] have been applied to predict age from structural MRI data. Recently, a study [5] compared eight ML-based models for predicting age and cognitive functions using diffusion MRI connectivity data. They found that a multilayer perceptron network (MLP) achieved the best results.

Although the above methods successfully predicted different phenotypic traits, they did not specifically investigate the multi-view nature of MRI data. In this paper, we investigate multi-view learning [6–9] to handle a specific type of multi-view data *where the same set of objects (samples) is described by several distinct feature sets [9]*. This type of data occurs naturally in dMRI, where various microstructure models can offer unique information (multiple views) [10] to describe each anatomical connection. Similarly, parcellated structural MRI data can contain multiple views of each anatomical structure, such as the thickness and surface area of each cortical region. For clarity, we call this type of data *anatomical-multi-view data*, where multiple feature sets describe each anatomical structure of the brain's WM or GM. Most related work in multi-view learning for neuroimaging has focused on multimodal images or sets of extracted image features from multimodal imagery [5,11–24]. Thus, there has been limited focus on the development of methods specifically for anatomical-multi-view data derived from MRI.

Challenges for learning from anatomical-multi-view data include: (1) challenge to leverage the underlying feature patterns of each anatomical view, (2) challenge to fuse information from other views while learning view-specific features, (3) challenge to simultaneously interpret each view and each structure's importance for different learning tasks, and (4) challenge to develop a general model that can perform well across different input datasets and tasks. To address these challenges, we propose an explainable multi-view network (EMV-Net). First, we apply an individual feature extractor with sparse self-attention to each view to extract its underlying patterns [17]. Second, we propose a method for cross-view calibration that can fuse features across views without losing view-specific information. Specifically, a wavelet transform that can decompose relevant information to reveal latent feature patterns in the frequency domain, is employed to improve the calibration. Third, our proposed network allows interpretable explanations of the importance of each anatomical view and different brain region. Specifically, the calibrator assigns learnable weights for each view before fusing them, which explain the importance of each view. The decomposed information from the wavelet transform in the calibrator is transformed into attention scores, which indicate the importance of anatomical regions. Finally, the network is evaluated using two different inputs (white matter and gray matter anatomy) and on three different prediction tasks. Overall, the network has superb performance, demonstrating that it can extract distinct information of a certain view from all anatomical structures while synthesizing multiple feature sets (views) for a specific structure.

## 2    Methodology

### 2.1    Dataset

**White Matter Feature Dataset.** The dataset of diffusion measure included in this paper is derived from the "1200 Subjects Data Release" dataset from the Human Connectome Project (HCP) Young Adult Study [25]. The data processing pipeline for tractography and tract parcellation is described in [26]. Briefly, to compute whole-brain tractography, the two-tensor unscented Kalman filter (UKF) [27] method is employed via the ukftractography package of SlicerDMRI [28]. Following a recursive estimation order, the UKF method fits a mixture model of two tensors to the diffusion data while tracking fibers. The first tensor models the traced tract, and the second tensor models fibers crossing the tract. UKF tractography is highly consistent across ages, health conditions, and image acquisitions [29,30]. Afterward, tractography parcellation is performed based on a neuroanatomist-curated WM atlas [30–32]. The tractography parcellation contains 947 clusters after discarding expert-defined false positive clusters as in [30].

For each cluster, we compute a total of 22 features, where each feature is considered as a view of the cluster. We compute the FA and Trace from Tensor 1 and Tensor 2 at each streamline point and record their max, min, median, mean, and variance across all fiber points within the cluster. Therefore, given each tensor $t \in \{t_1, t_2\}$, we extract statistical measures $s \in \{max, min, median, mean, var\}$ of the diffusion parameters $p \in \{FA, Trace\}$, resulting in 20 features per cluster. We also compute the total number of points and streamlines in each cluster.

**Gray Matter Feature Dataset.** FreeSurfer [33] is a widely-used tool for automated segmentation and parcellation of brain regions from MRI. The dataset used in this study was generated by the Freesurfer 5.3.0-HCP pipeline and contains measurements of 68 cortical regions. For each region, the dataset offers two sets of features: the surface area that reflects the number and density of neurons in the anatomical region and the surface thickness that relates to the integrity of cortical tissue [34]. These features are considered two anatomical views in our study.

**Predicting Tasks.** We choose three popular brain-based prediction tasks: age [4] (demographic), strength [35] (motor), and picture vocabulary [5] (cognition).

### 2.2    Methods

The overall architecture of the proposed EMV-Net is shown in Fig. 1, which is used to extract view-specific features, and the cross-view calibrator (Fig. 2) that is proposed fuse features from multiple views to calibrate view-specific information and provide interpretability for each view and anatomical region's importance.

**Single-view Backbone.** For each view, we apply a feature extractor with a commonly used backbone, largely following the design of [36]. In brief, the backbone has three repeated modules (placed sequentially), where each module utilizes a patch embedding unit for downsampling, then extracts information with multiple sparse CMT (convolution-meets-transformer) blocks (S-CMT Blocks). By combining convolution and self-attention outputs, the backbone can extract both local and global information. Note that unlike [36], we use 1.5-Entmax [37] for self-attention, which can sparsely select the most relevant anatomical structures from each view.

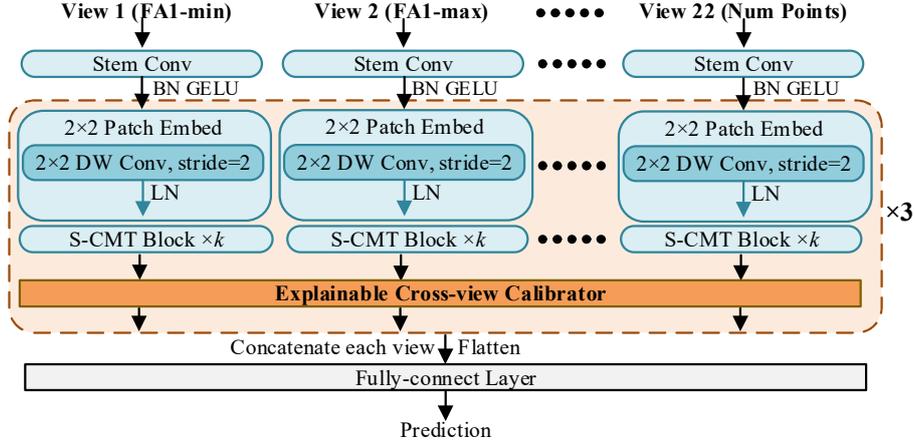

**Fig. 1.** The overall architecture of the proposed explainable attentive multiview network.

**Attentive Multiview Network.** As shown in Fig. 1, we employ a multi-view-multi-net architecture like [11]. However, instead of only performing one fusion as in [11], we fuse information across views in each repeated module using the explainable cross-view calibrator, which enables view-specific feature extractors to learn complementary and relevant information from other views. After the last calibrator, feature maps from all views are concatenated, flattened, and then used to perform predictions. The overall architecture of the cross-view calibrator is shown in Fig. 2. It contains three steps: (1) highlighting relevant features from multiple views and fusing them, (2) performing wavelet transform on the fused features, and (3) sigmoid normalization to produce attention scores from the processed information and then applying these scores to calibrate view-specific features.

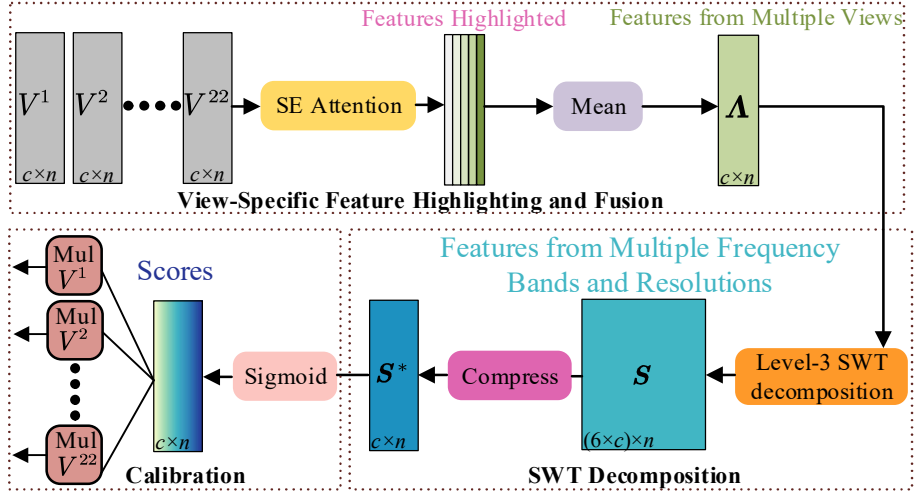

**Fig. 2.** The proposed explainable cross-view calibrator. $n$ and $c$ denote the feature length (number of features from all clusters) and the number of channels.

*View-specific feature highlighting and fusion.* Since different anatomical views can have different contributions to different tasks, applying flexible weights to the views could boost performance while granting interpretability. To highlight more relevant views, we first apply a squeeze-and-excitation module [38] to produce attention weights for each view (here we treat each view as a channel) and assign these weights to the corresponding view features. The selected features from each view are then fused by averaging, which produces the fused feature map $\Lambda$.

*SWT decomposition for improved calibration across views:* The wavelet transform [39] is a popular signal processing tool and has been applied to denoise or extract features from structural and diffusion MRI [40–42]. It can reveal the latent patterns of a signal by decomposing it into low- and high-frequency bands and achieve multi-resolution analysis by performing further decomposition over the low-frequency band. For our data, the wavelet transform could obtain the rough pattern of features for anatomical structures from the low-frequency band, and fine details from the high-frequency bands. Rather than transforming our input data directly, we propose to use this frequency information to improve calibration across views (see calibration details below). The stationary wavelet transform (SWT) [43] is an improved version of the discrete wavelet transform (DWT) and has different decomposition strategies. As shown in Fig. 3, SWT upsamples its filters by a factor of $2^{(i-1)}$ at the $i$-th level, which allows the output to retain the original resolution and avoid aliasing artifacts due to downsampling of DWT [44]. In addition, the denoising property of SWT also helps to remove each view's irrelevant information. However, the output of SWT could be "redundant" due to the absence of downsampling. To mitigate this, we add a linear projection layer to compress the decomposed features from SWT. As each level of SWT generates a low- and a high-frequency subband that is of the same size as $\Lambda$, the output of a 3-level SWT is $S \in \mathfrak{R}^{(6 \times c) \times n}$. Note that we concatenate each level's output according to the channel. Then, a linear projection is performed upon channel dimension to compress $S$ to $S^* \in \mathfrak{R}^{(6 \times c) \times n}$. This facilitates the fusion of low- and high-frequency features from different scales.

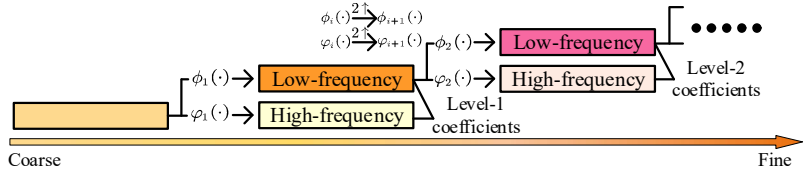

**Fig. 3.** The process of SWT decomposition. Here the low-pass and high-pass filters $\phi(\cdot)$ and $\varphi(\cdot)$ are upsampled after each level of decomposition.

*Calibration:* Since each anatomical view could provide unique information and could be important for prediction, instead of directly applying the decomposed information from SWT for each view as the new input, we apply it to calibrate each view's original information. Motivated by [45] that produced attention from the discrete cosine transform, we apply Sigmoid to process the frequency information from $S^*$. The resultant score $AS$ is then applied to calibrate each view-specific information $V^i$, as in the below equation.

$$AS = Sigmoid(S^*), \ V^i = V^i \times AS \times \gamma + \beta$$

where $\gamma$ and $\beta$ are two learnable parameters incorporated to ease optimization [46]. The produced $AS$ also provides interpretability that explains which anatomical region (the feature dimension of $V^i$) has higher contribution to the calibration.

## 3   Experiments

### 3.1   Experimental Settings

We evaluate the proposed model using Pytorch 1.12.1 and Nvidia RTX3090 card. All tests are based on 10-fold cross-validation. Before model training, L-2 normalization is applied to project the features into the same scale. The optimizer we choose is AdamW, with a weight decay of 0.05 and a learning rate of 0.01. To facilitate the convergence, we employ the cosine annealing learning rate scheduler. The loss is mean-squared loss. For SWT, we choose bior2.4 as the mother wavelet.

### 3.2   State-of-the-art and Baseline Comparison

We choose the single-view backbone (Single-view) as a baseline and train it by concatenating all views and treating the concatenated features as one view. Furthermore, we remove the cross-view calibrator from EMV-Net (w/o cross-view calibrator) to show its effectiveness. Additionally, we add three ML-based models that were pro-posed for brain-based prediction, including SFCN [4], MLP [5], and ridge regression [3]. Here we include the WM dataset and the GM dataset and test models' performances on age (demographic), strength (motor), and picture vocabulary (cognition) prediction. The results are shown in Table 1. Note that the metric for age prediction is the mean absolute error (MAE) and the metric for the other two is the correlation coefficient, which are popular metrics for the three tasks [4,5,35]. We further perform the repeated-measure ANOVA test and then the paired t-test to demonstrate the significance of the proposed EMV-Net when compared with other baselines and methods, as in Table 1.

**Table 1.** Compare EMV-Net with and without the cross-view calibrator (in light gray). Also, compare the single-view baseline (in light gray) and three state-of-the-art methods in brain-based prediction (in dark gray). Note that the metric for age prediction is MAE, and the metric for strength and picture vocabulary prediction is the Spearman correlation coefficient. The paired t-test results for comparative implementations against the proposed one (EMV-Net) are presented by asterisks. * indicates that $p<0.05$, and ** indicates that $p<0.001$.

| | **EMV-Net** | | | w/o Cross-view Calibrator | | | Single-view | | |
|---|---|---|---|---|---|---|---|---|---|
| | Age | Strength | PicVocab | Age | Strength | PicVocab | Age | Strength | PicVocab |
| WM | 2.51 | 0.69 | 0.38 | 2.61** | 0.66* | 0.36* | 2.87** | 0.60** | 0.35* |
| GM | 2.80 | 0.56 | 0.39 | 2.86* | 0.55 | 0.38 | 3.03** | 0.53* | 0.36* |
| | SFCN | | | MLP | | | Ridge Regression | | |
| | Age | Strength | PicVocab | Age | Strength | PicVocab | Age | Strength | PicVocab |
| WM | 2.75** | 0.56** | 0.35* | 2.98** | 0.53** | 0.19** | 2.90** | 0.61** | 0.31** |
| GM | 2.94** | 0.54 | 0.37 | 3.06** | 0.30** | 0.25** | 2.83* | 0.52* | 0.28** |

### 3.3 Ablation Study

As presented in Table 2, we further show that the self-attention with 1.5-Entmax could outperform other activations (Softmax and Sparsemax). In addition, we compare different designs of the cross-view calibrator on whether SWT is applied (No SWT) or other levels of SWT is applied (level 1 and 2). All comparisons are based on the WM dataset. Based on the results, we test the statistical significance of our design. For the two groups in Table 2, repeated-measure ANOVA tests indicate significant differences. Furthermore, we do paired t-tests and demonstrate that our design significantly outperforms alternatives.

**Table 2.** Ablation studies over the (1) activation functions for self-attention (compared to using 1.5-Entmax). (2) SWT for the cross-view calibrator. The paired t-test results for comparative implementations against the proposed one are presented by asterisks. * indicates that $p<0.05$, and ** indicates that $p<0.001$. The metrics for each task are the same as in Table 1.

| | | (1) | | (2) | | |
|---|---|---|---|---|---|---|
| | **Proposed** | Sparsemax | Softmax | No SWT | SWT Level 1 | SWT Level 2 |
| Age | **2.51** | 2.54* | 2.73** | 2.65* | 2.58* | 2.56* |
| Strength | **0.69** | 0.68 | 0.65* | 0.60* | 0.68 | 0.69 |
| PicVocab | **0.38** | 0.35* | 0.28** | 0.30** | 0.36 | 0.36 |

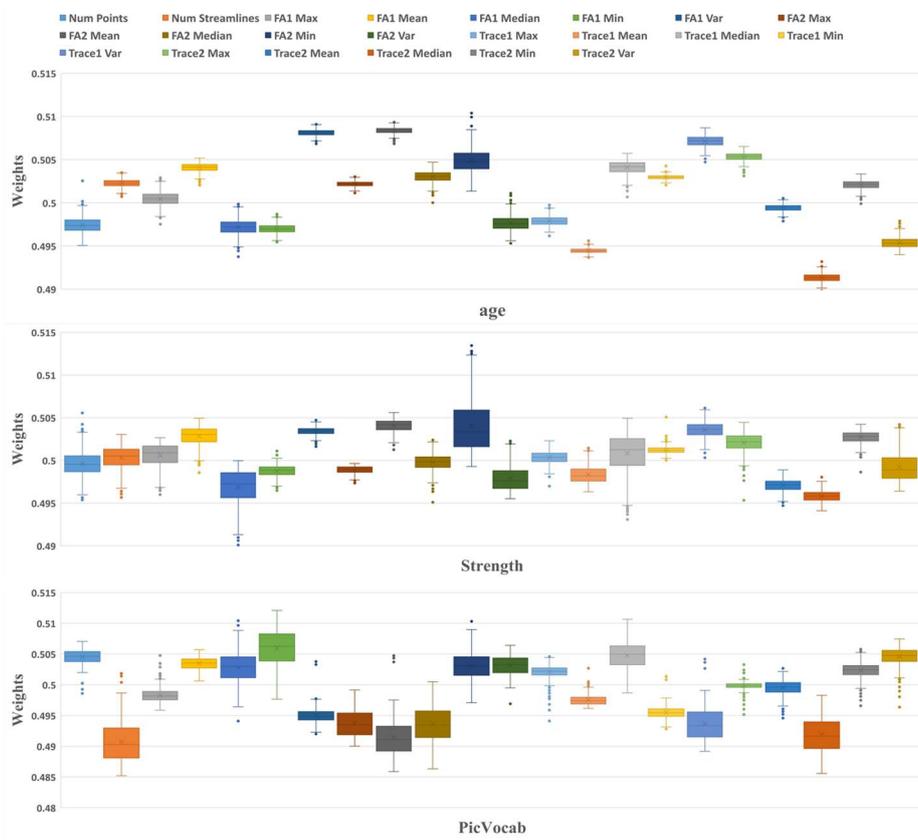

**Fig. 4.** Normalized weights for 22 anatomical views based on the WM dataset.

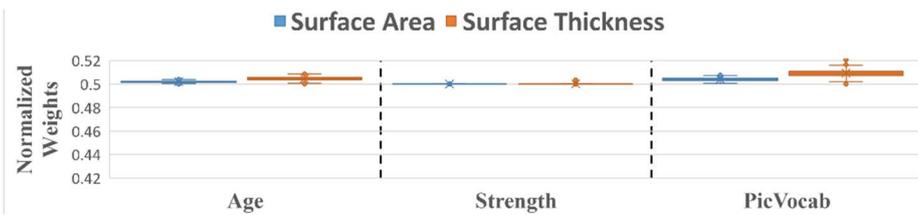

**Fig. 5.** The normalized weights for the 2 anatomical views based on the GM dataset.

### 3.4 Interpretability Analysis

**Compare Different Anatomical Views**: We further analyze each anatomical view's importance for the three prediction tasks. We calculate each anatomical view's attention weights that produced from the view-specific feature highlighting phase of three cross-view calibrators based on the WM and GM feature datasets. Each view contains 1065 weights for the 1065 subjects. We average the weights and present the results as box plots.

From Fig. 4, the three tasks have different weights for different views, which means that the model focus on different dMRI measures for prediction. For age prediction based on the WM dataset, FA1 variance and FA2 mean play the most important role, while Trace2 median is the least important. Other views such as Trace1 variance and Trace2 max are also important for age prediction. For strength prediction, the four views above are also pivotal. Apart from them, the model also focuses on FA1 mean and FA2 min, while paying the least attention to Trace2 mean and Trace2 median. For picture vocabulary prediction, FA1 min contributes the most, while it is less important for the other two tasks. In addition, the number of streamlines is the least important for picture vocabulary prediction, whereas this view is relatively more significant for the other two tasks. Furthermore, while the model for age and strength prediction assigns higher weights to FA1 variance and FA2 mean, picture vocabulary prediction largely overlooks these two. It should be noted that the strength prediction has a more balance focus on the 22 views than the other two tasks (the variance of weights for the 22 views of strength prediction is less than the other two).

The results on prediction based on the GM data is shown in Fig. 5. From the figure, surface thickness is more important for age and picture vocabulary prediction. For strength prediction, both anatomical views are almost equal in importance.

## 4 Conclusion

In this study, we investigated a specific type of multi-view data that commonly occurs in segmented or parcellated MRI, where each anatomical structure is described with multiple features. We proposed to call this type of data anatomical-multi-view data. To efficiently extract useful features from multiple views while preserving interpretability, we have proposed an explainable multi-view network to learn both view-specific and across-view information. The model performed well on three different prediction tasks using WM and GM datasets. We further presented interpretable analyses of the importance of different anatomical views over the three tasks. We found that the FA1 variance and FA2 mean are important for age and strength prediction, while they are trivial to picture vocabulary prediction. On the contrary, FA1 min is pivotal to picture vocabulary prediction, whereas it is less important for the other two tasks. Apart from this, for prediction based on the GM dataset, we found that surface thickness is slightly more important than surface area for age and picture vocabulary prediction, while these two are almost equal in significance for strength prediction. Overall, this investigation suggests that the exploration of methods designed specifi-cally for anatomical-multi-view data holds potential for the study of the brain using machine learning.

## 5    Declaration of Conflict

There are no financial or non-financial conflict of interest.